\documentclass[a4paper]{jpconf}
\usepackage{graphicx}
\usepackage{caption}
\usepackage{subcaption}
\usepackage{lineno}
\begin{document}
\title{(Anti-)(hyper-)nuclei production in Pb--Pb collisions with ALICE at the Large Hadron Collider}
%\linenumbers
\author{Stefano Trogolo for the ALICE Collaboration}

\address{Universit\`a degli Studi di Padova and INFN - Sezione di Padova \\ Via Marzolo, 8 - 35131 Padova, Italy 
}

\ead{stefano.trogolo@cern.ch}

\begin{abstract}
%The ALICE Collaboration has collected a large data sample of Pb-Pb collisions at $\sqrt{s_{\rm{NN}}}$ = 5.02 TeV in 2015 with the Large Hadron Collider (LHC). 
In ultra-relativistic heavy-ion collisions a great variety of (anti-)(hyper-)nuclei are produced, namely deuteron, triton, $^3$He, $^4$He, hypertriton ($^3_{\Lambda}$H) and their antiparticles. The ALICE experiment is the most suited to investigate the production of (hyper-)nuclei at the LHC, thanks to an excellent particle identification and low-material budget detectors.
Recent results on (hyper-)nuclei production as a function of transverse momentum ($p_{\rm{T}}$) and charged particle multiplicity (d$N_{\rm{ch}}$/d$\eta$) in Pb--Pb collisions are presented. The evolution of the production yields with the system size is also shown, via comparison to the results obtained in small collision systems, like pp and p--Pb.
The results on the production of (hyper-)nuclei are also compared with the predictions based on a naive coalescence approach and the statistical hadronization models

Furthermore, the latest and currently most precise measurement of the hypertriton lifetime is presented. It is compared with results obtained by different experimental techniques and with theoretical predictions. 

\end{abstract}

\section{Introduction}
In ultra-relativistic heavy-ion collisions at the LHC high-energy density and temperature are reached. In this regime the colour-charge deconfinment is achieved and a state of matter called Quark-Gluon Plasma (QGP) is created. Among the particles produced in these collisions, (hyper-)nuclei and their antiparticles are of special interest since the production mechanism of these loosely bound ($E_{B} \sim$ 1 MeV) states is not clear.
A hypernucleus is a nucleus where one nucleon is replaced by a hyperon. The hypertriton ($^3_{\Lambda}$H), the lightest known hypernucleus, is a bound state of a proton, a neutron, and a $\Lambda$ baryon. 

There are two classes of phenomenological models used to describe the production of \mbox{(anti-)(hyper-)nuclei} in heavy-ion collisions: the statistical-thermal model \cite{Thermal} and the coalescence of baryons \cite{Coal_1, Coal_2}.
According to the Statical Hadronisation Model (SHM), (hyper-)nuclei and hadrons are produced from the created medium, in statistical equilibrium, at the chemical \mbox{freeze-out}, when inelastic collisions cease. This model can describe the production yields (d\textit{N}/d\textit{y}) in Pb--Pb collisions rather well using a grand canonical approach, where two key parameters are the source volume V and the chemical freeze-out temperature $T_{\rm{chem}}$.

The coalescence approach assumes that (anti-)baryons close enough in the phase-space at kinetic freeze-out, when elastic collisions cease, can form a multi-baryon state. The key observable is the coalescence parameter, defined as:
\begin{equation}
	B_A = E_A\frac{\mathrm{d}^{3}N_A}{\mathrm{d}p^{3}_{A}} \left( E_p\frac{\mathrm{d}^3N_p}{\mathrm{d}p^3_p}\right)^{-A}
\end{equation}
where $E_A\frac{\mathrm{d}^3N_A}{\mathrm{d}p^3_A}$ and $E_p\frac{\mathrm{d}^3N_p}{\mathrm{d}p^3_p}$ are the invariant production spectra of nucleus of mass A and proton, respectively.

The study of (anti-)(hyper-)nuclei production in heavy-ion collisions is fundamental to improve our knowledge on their production mechanism and to constrain theoretical models. An overview of the results obtained by analysing the data sample of Pb--Pb collisions at \mbox{$\sqrt{s_{\mathrm{NN}}}$ = 5.02 TeV} is reported in this proceedings.

Furthermore, the $^3_{\Lambda}$H lifetime is one of the open points of  the hypernuclear physics. According to theoretical predictions it should be slightly below the free $\Lambda$ baryon lifetime, but recent and precise measurements showed a trend away from the expected value. A new and more precise lifetime measurement is reported in these proceedings.

\section{A Large Ion Collider Experiment}
The ALICE experiment \cite{Abelev:2014ffa} has excellent particle identification (PID) capabilities which allow for the detection of these rarely produced particles. PID is performed by exploiting the measurement of the specific energy-loss in the Time Projection Chamber (TPC) and the information of the Time-Of-Flight (TOF) detector. 
In addition, the high-resolution vertexing provided by the Inner Tracking System (ITS) is used to distinguish primary from secondary vertices originating from weak decays, as in the analysis of the (anti-)hypertriton which has a decay length of several centimeters. The latter is reconstructed via the 2-body decay mode, $^3_{\Lambda}$H $\rightarrow ^3$He + $\pi^-$, which is the one detected with the highest reconstruction efficiency, even though the 3-body decay channel has the highest branching ratio (B.R.) $\sim$ 40$\%$.

\begin{figure}[!h]
     \centering
     \begin{subfigure}[b]{0.49\textwidth}
         \centering
         \includegraphics[width=0.965\textwidth]{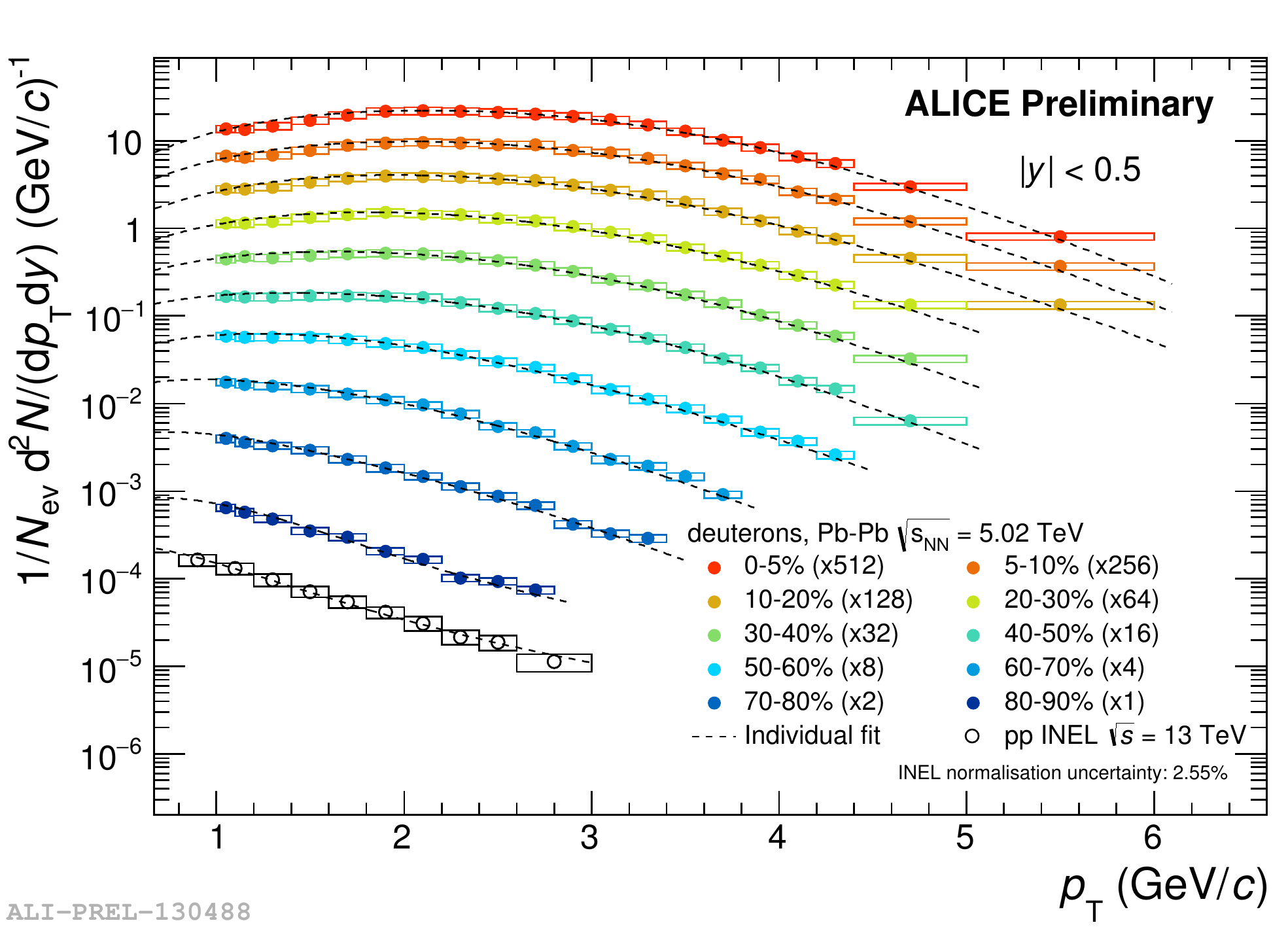}
         \caption{ }
         \label{deu_spectrum}
     \end{subfigure}
     \hfill
     \begin{subfigure}[b]{0.49\textwidth}
         \centering
         \includegraphics[width=\textwidth]{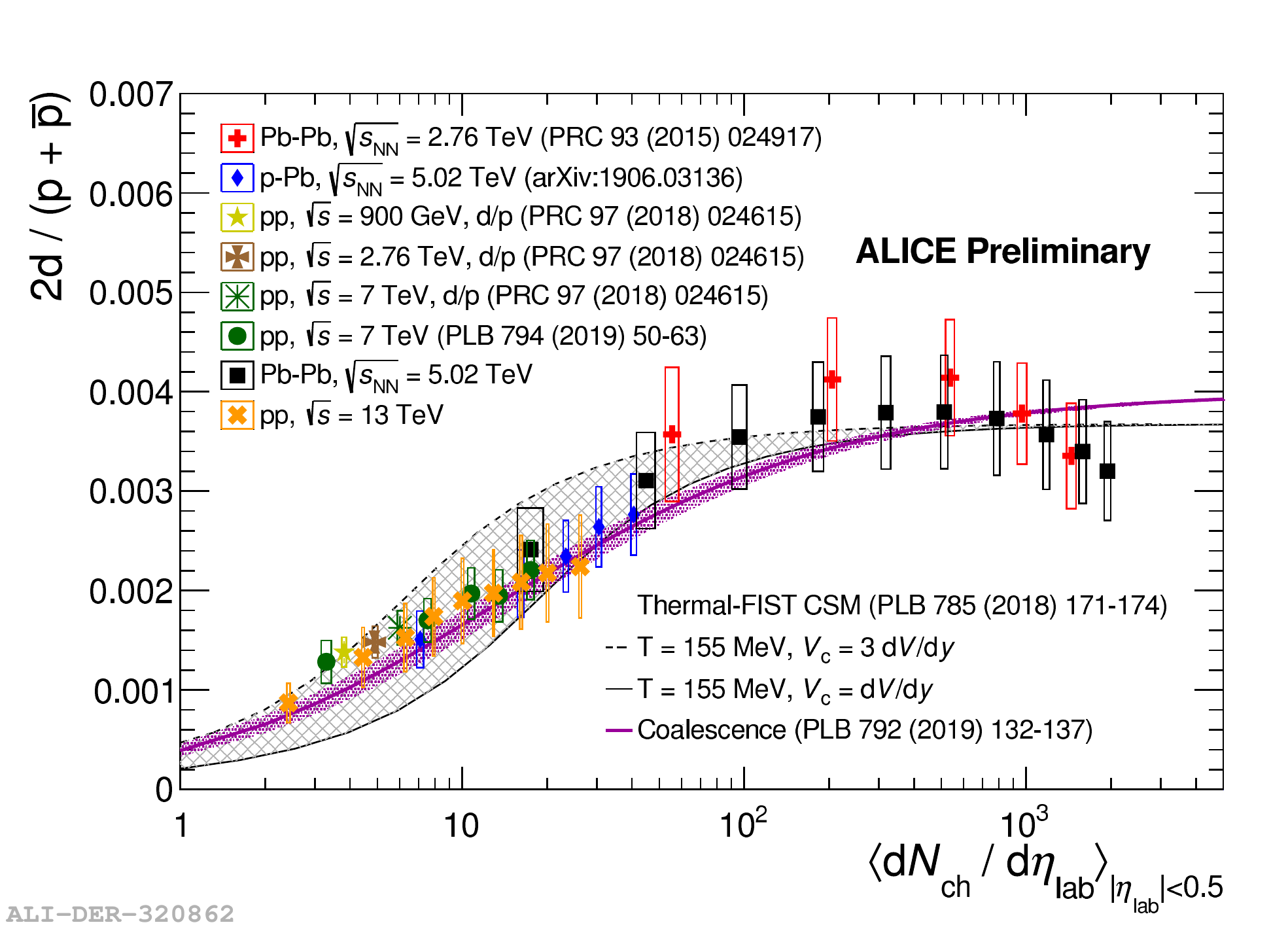}
         \caption{ }
         \label{doverp}
     \end{subfigure}
        \caption{(a) $p_{\mathrm{T}}$-differential production spectra of deuterons measured in different centrality classes of Pb--Pb collisions at $\sqrt{s_{\mathrm{NN}}}$ = 5.02 TeV. The distribution with open markers is the deuteron $p_{\mathrm{T}}$ spectrum measured in pp collisions at $\sqrt{s}$ = 13 TeV; (b) deuteron-over-proton ratio of the $p_{\mathrm{T}}$-integrated yields, for different collision systems, as a function of d$N_{\rm{ch}}$/d$\eta$. The ratios are compared with the predictions of the coalescence model \cite{Sun:2018mqq} and Thermal-FIST CSM \cite{Thermal_fist}.}
        \label{fig_1}
\end{figure}

\section{Nuclei production}
The ALICE experiment has measured the $p_{\mathrm{T}}$-differential production spectra for \mbox{(anti-)deuterons} and \mbox{(anti-)$^3\mathrm{He}$} in multiple centrality classes, corresponding to different charged particle multiplicities, of Pb--Pb collisions. 

Figure~\ref{deu_spectrum} shows the deuteron production spectra in Pb--Pb collisions at $\sqrt{s_{\mathrm{NN}}}$ = 5.02 TeV, for different centrality classes, and in pp at $\sqrt{s}$ = 13 TeV \cite{ALICE-PUBLIC-2017-006}. The spectra  exhibit a hardening with increasing collision centrality in Pb--Pb, which can be explained by hydrodynamic models as an effect of the radial flow. Similar behaviour is observed also for (anti-)$^3\mathrm{He}$. In order to extrapolate the spectra to the unmeasured $p_{\mathrm{T}}$ regions and to calculate the integrated yields (d$N$/d$y$), the distributions have been fitted with the Blast-Wave \cite{BlastWave} function.

In Fig.~\ref{doverp} the ratio between the $p_{\mathrm{T}}$-integrated yields of deuterons and protons, measured in Pb--Pb collisions, is shown together with the ratio obtained in smaller collision systems at different energies, namely pp and p--Pb. The data are also compared with the prediction of coalescence calculations \cite{Sun:2018mqq} and of the canonical statistical model (CSM) using \mbox{Thermal-FIST \cite{Thermal_fist}}. A smooth evolution of the ratio with the event multiplicity, hence with the system size, is visible. The rise at low multiplicity can be interpreted in the coalescence approach as related to the small phase-space of the nucleons and for the statistical model it is the result of canonical suppression. The flat trend at high multiplicity, typical of Pb--Pb collisions, is described by both models. The same behaviour is observed also in the $^3\mathrm{He}$/p, leading to the same interpretation.

\begin{figure}[!t]
     \centering
     \begin{subfigure}[b]{0.49\textwidth}
         \centering
         \includegraphics[width=0.965\textwidth]{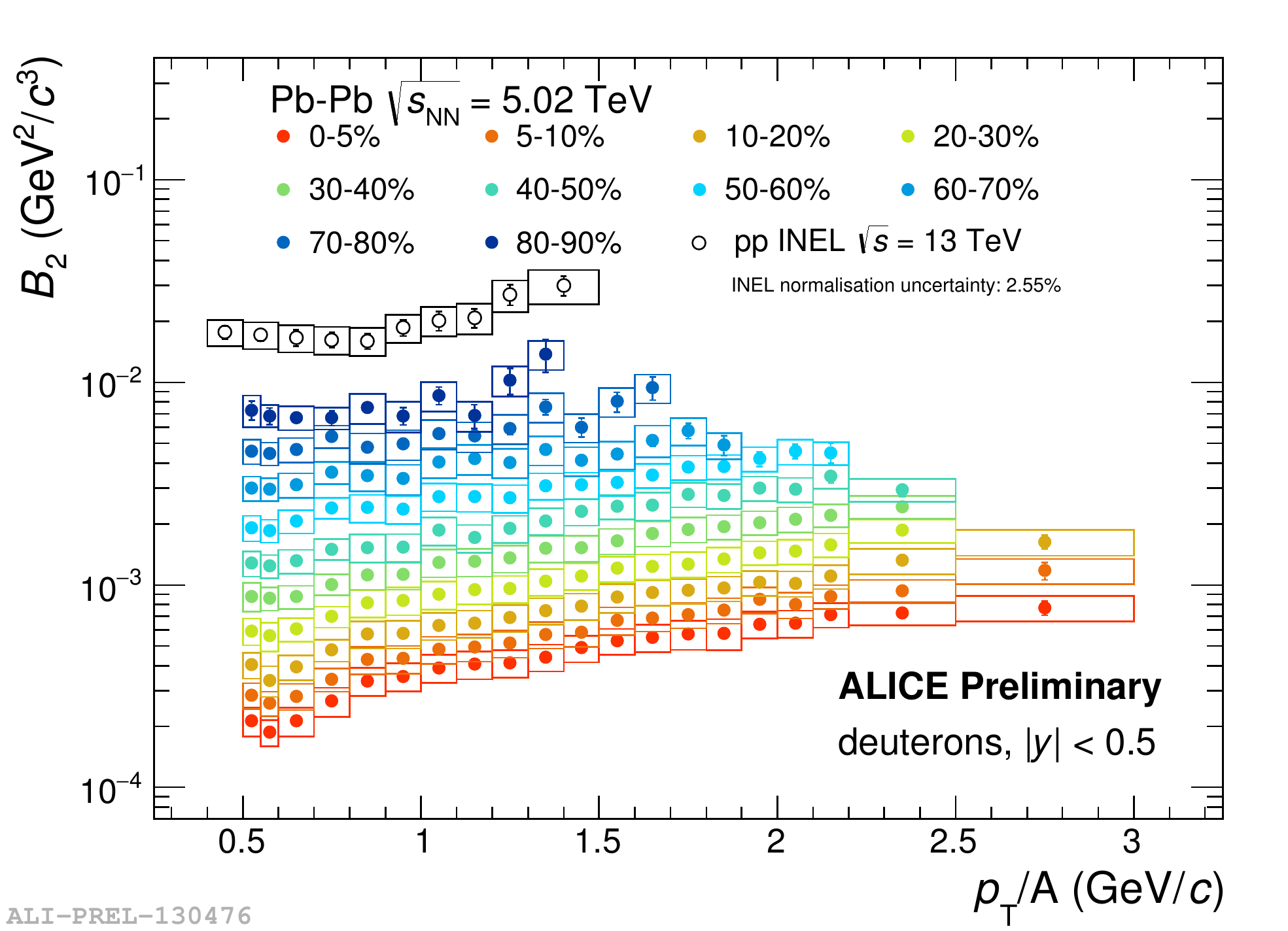}
         \caption{ }
         \label{b2_deu}
     \end{subfigure}
     \hfill
     \begin{subfigure}[b]{0.49\textwidth}
         \centering
         \includegraphics[width=\textwidth]{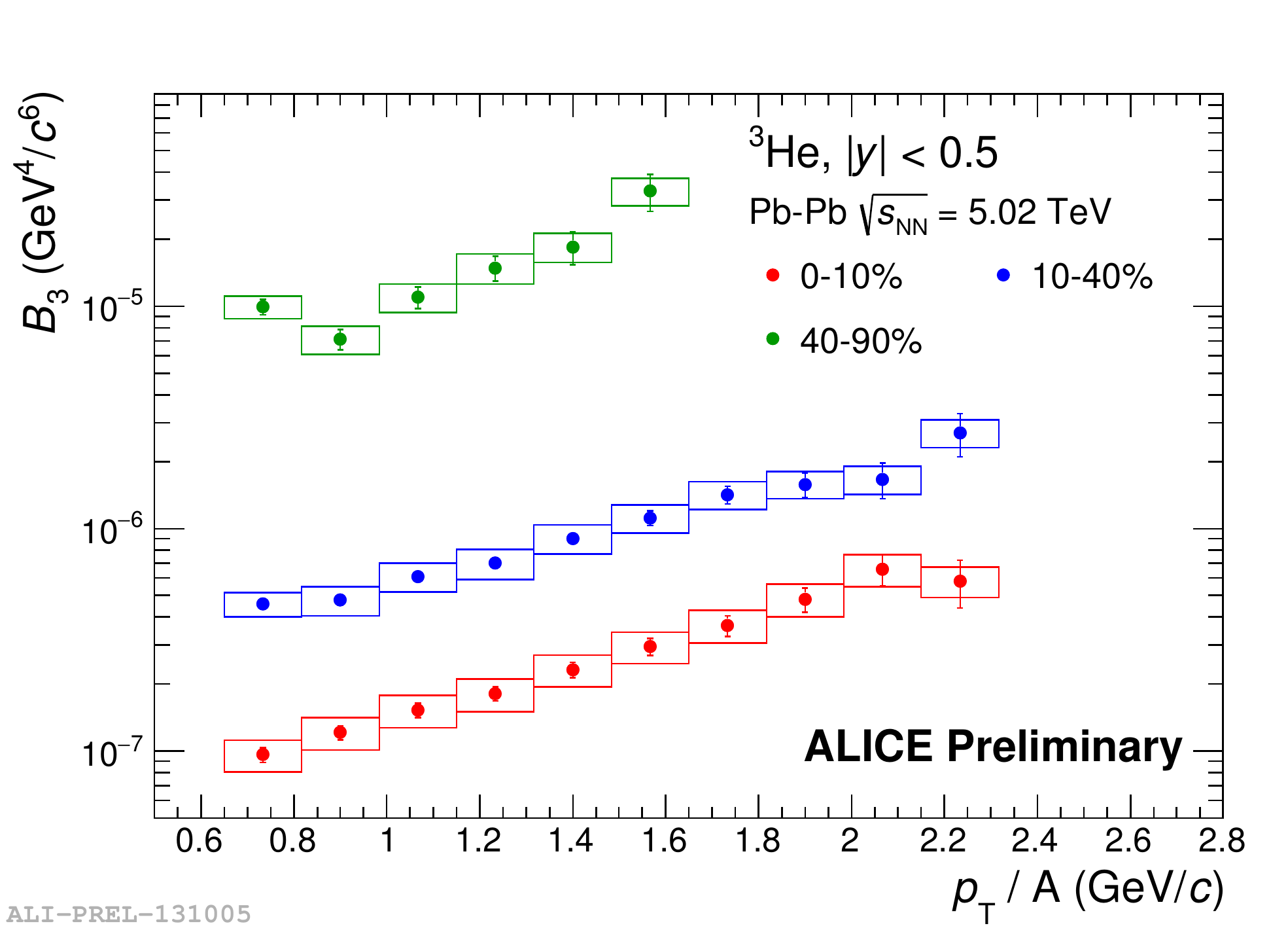}
         \caption{ }
         \label{b3_he3}
     \end{subfigure}
        \caption{Coalescence parameter $B_{\mathrm{2}}$ of deuterons (a) and $B_{\mathrm{3}}$ of $^3\mathrm{He}$ (b) as a function of $p_{\mathrm{T}}$/A for different centrality classes in Pb--Pb at $\sqrt{s_{\mathrm{NN}}}$ = 5.02 TeV.}
        \label{fig_2}
\end{figure}

Figure~\ref{fig_2} shows the coalescence parameter for deuteron (a), $B_{\mathrm{2}}$, and $^3\mathrm{He}$ (b), $B_{\mathrm{3}}$, measured in different centrality classes of Pb--Pb collisions. According to the naive coalescence model predictions $B_{\mathrm{A}}$ should be flat as a function of $p_{\mathrm{T}}$/A. However, the observed trend is not in agreement with simple coalescence and more refined approaches \cite{heinz1999,blum2017,zhao2018}, taking into account the spatial extension of the source can be helpful for a meaningful comparison. On the other hand, the ordering of $B_{\mathrm{A}}$ with centrality is explained in the coalescence model as related to the increasing size of the emitting source volume.

\section{Hypertriton production}
The (anti-)hypertriton production and lifetime are measured by reconstructing the decay mode into two bodies, $^3_{\Lambda}$H $\rightarrow ^3$He + $\pi^-$, by an invariant mass analysis.

\begin{figure}[!h]
     \centering
     \begin{subfigure}[b]{0.49\textwidth}
         \centering
         \includegraphics[width=1.1\textwidth]{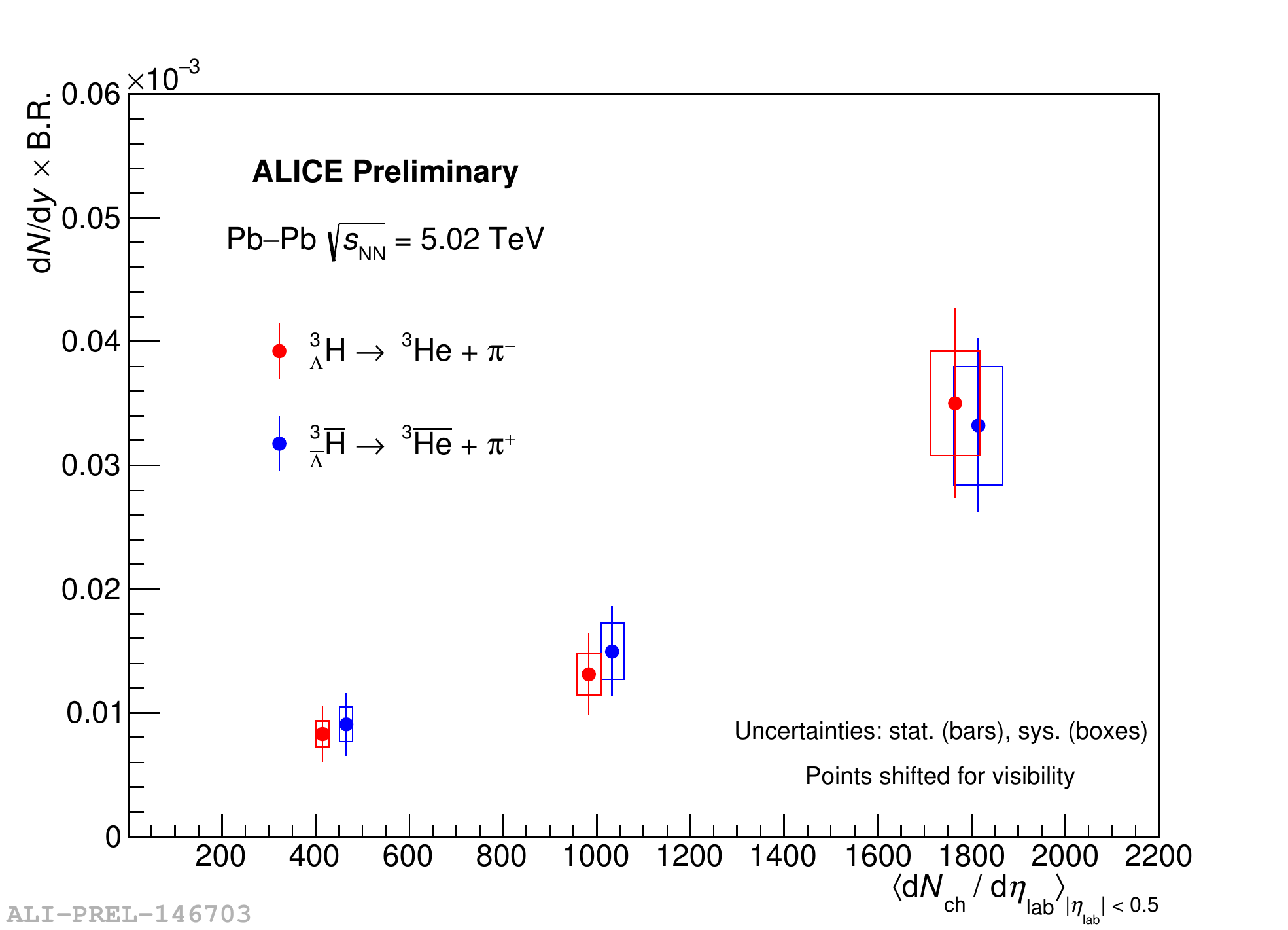}
         \caption{ }
         \label{3lh_mult}
     \end{subfigure}
     \hfill
     \begin{subfigure}[b]{0.49\textwidth}
         \centering
         \includegraphics[width=0.82\textwidth]{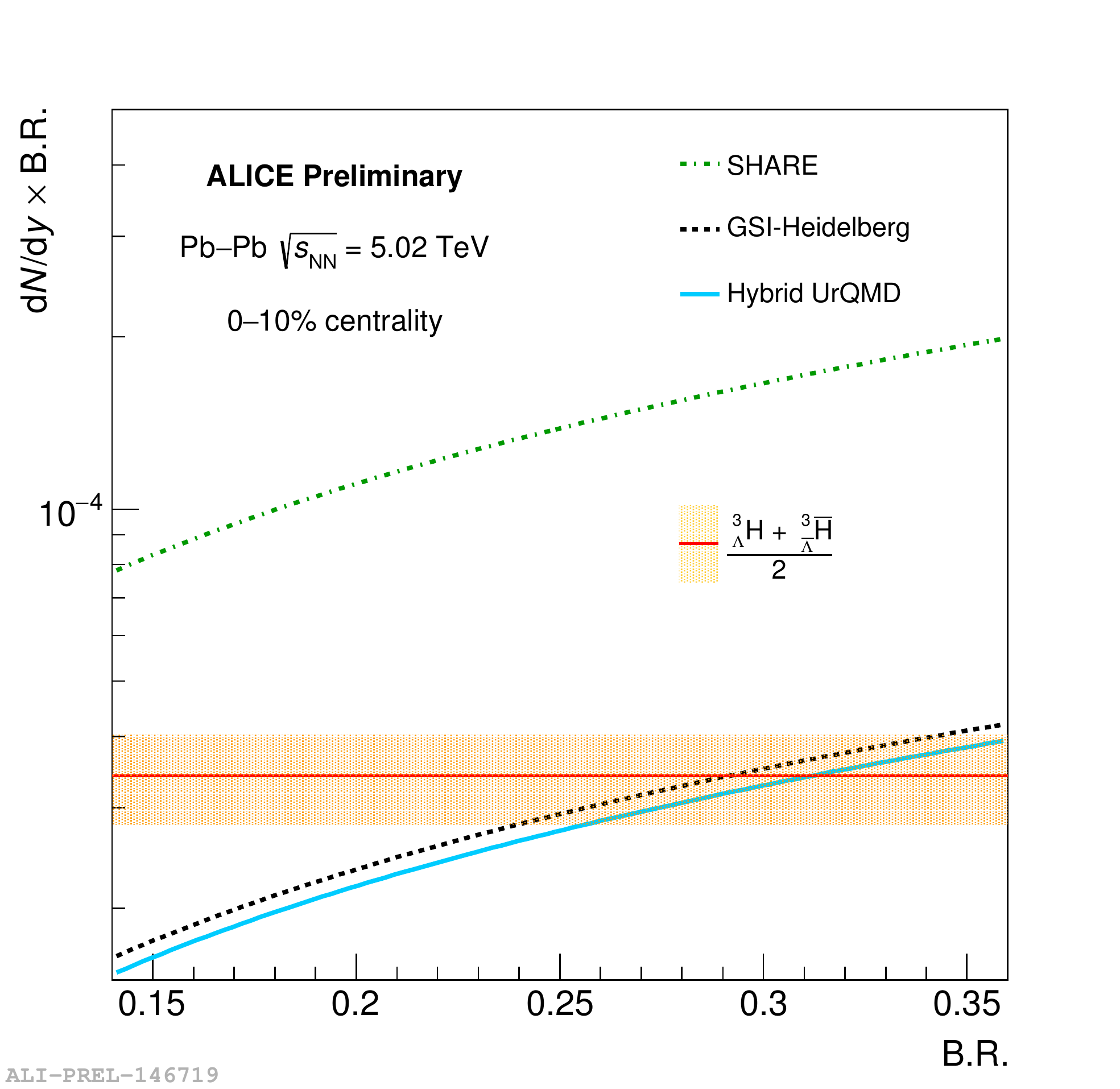}
         \caption{ }
         \label{3lh_br}
     \end{subfigure}
        \caption{(a) d$N$/d$y$ of $^3_{\Lambda}$H (red)and $^3_{\overline{\Lambda}}\overline{\mathrm{H}}$ (blue) as a function of the charge particle multiplicity d$N_{\rm{ch}}$/d$\eta$. Blue points are shifted for visibility; (b) averaged d$N$/d$y$ measured in central (0--10$\%$) Pb--Pb collisions compared with thermal model predictions \cite{Thermal,hyb_urqmd,share} as a function of the decay branching ratio (B.R.).}
        \label{fig_3}
\end{figure}

The $p_{\mathrm{T}}$-integrated yields d$N$/d$y$ have been measured in three different centrality classes, namely 0--10$\%$, 10--30$\%$, and 30--50$\%$. The corrected yields are shown in Fig.~\ref{3lh_mult} as a function of the charged particle multiplicity separately for $^3_{\Lambda}$H (red circle) and $^3_{\overline{\Lambda}}\overline{\mathrm{H}}$ (blue circle). According to the statistical hadronization models, the rising trend can be interpreted as a consequence of the increasing size of the medium.

The B.R. of the hypertriton, which is a key parameter, is barely known and mainly constrained by the ratio between all charged channels containing a pion, R$_{\rm{3}}$ = $\Gamma^{^{3}\mathrm{He}}$/($\Gamma^{^{3}\mathrm{He}}$ + $\Gamma^{p+d}$ + $\Gamma^{p+p+n}$) \cite{kamada1998}. Hence the d$N$/d$y$ measured in the 0--10$\%$ centrality class is compared with thermal model predictions as a function of the B.R. as shown in Fig.~\ref{3lh_br}. The  thermal models that assume equilibrium, GSI-Heidelberg \cite{Thermal} and Hybrid UrQMD \cite{hyb_urqmd}, describe the measured yield within the B.R. range 24-35$\%$, while the non-equilibrium model SHARE \cite{share} overestimates the yield by a factor 4 for the theoretically favoured B.R. of 25$\%$ \cite{kamada1998}.

\section{Thermal fit}
\begin{figure}[h]
	\centering
    \includegraphics[width=0.8\textwidth]{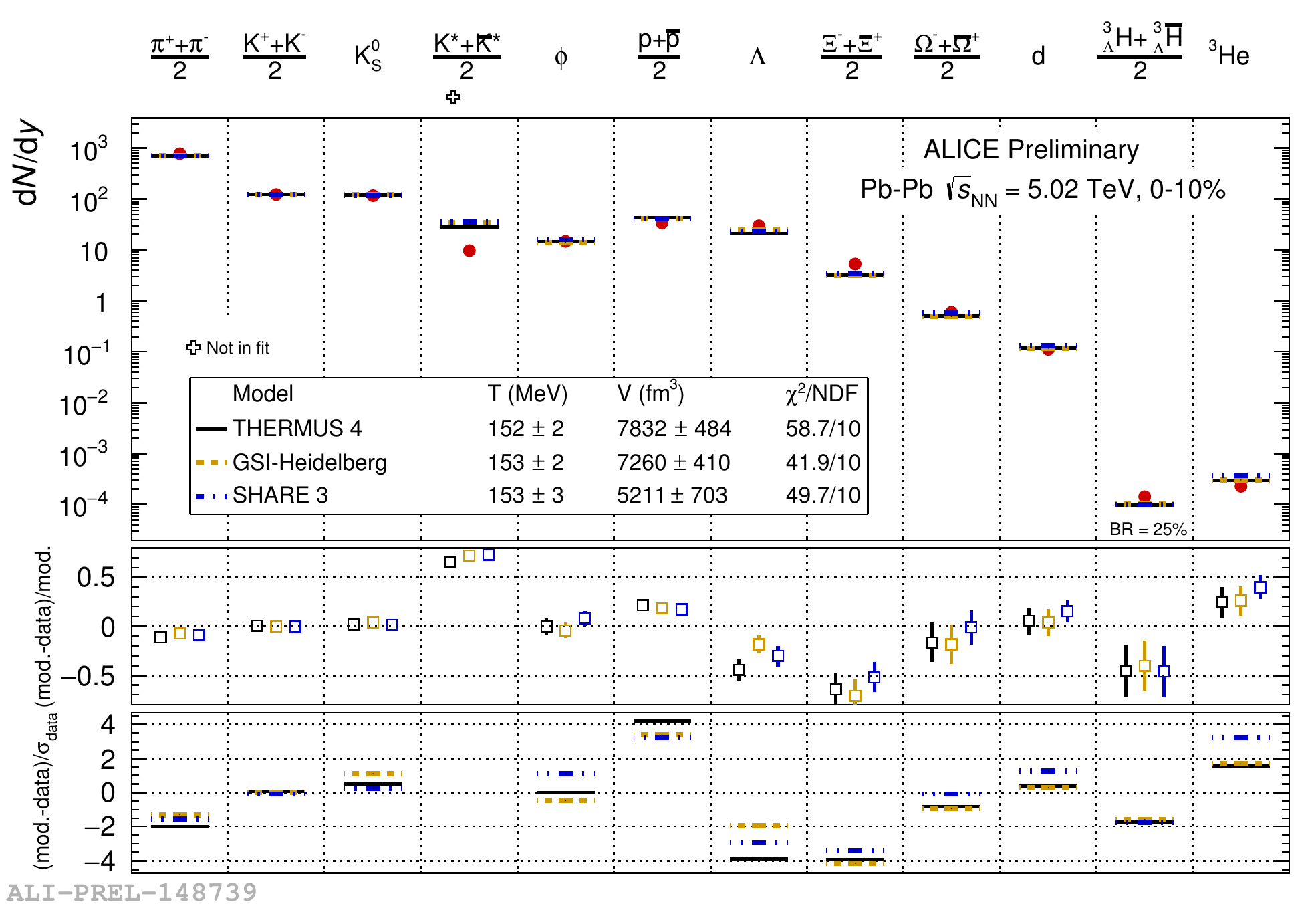}
	\caption{Thermal model fit with three different implementations \cite{Thermal,thermus,share} to the particle yields in central (0--10$\%$) Pb--Pb collisions at $\sqrt{s_{\mathrm{NN}}}$ = 5.02 TeV.}
    \label{thermal_fit}
\end{figure}

The nuclei and hypernuclei d$N$/d$y$, previously introduced, are included in the so-called thermal fit to all hadron production yields measured by the ALICE Collaboration in Pb--Pb collisions, as shown in Fig.~\ref{thermal_fit}. %$\sqrt{s_{\mathrm{NN}}}$ = 5.02 TeV in 0--10$\%$ centrality class
In particular three different implementations of the thermal model have been used, namely GSI-Heidelberg \cite{Thermal}, THERMUS \cite{thermus}, and SHARE \cite{share}. The d$N$/d$y$ are qualitatively well described by these models with $T_{\rm{chem}}$ = 153 $\pm$ 2 MeV and assuming statistical-thermal equilibrium. Nevertheless, the improved precision on d$N$/d$y$ measurements reached by ALICE leads to a tension between the data and the model. Hence, a further tuning and improvement of the models are needed.

\section{Hypertriton lifetime}
The $^3_{\Lambda}$H is a loosely bound system and in particular the $\Lambda$ separation energy is \mbox{$B_{\Lambda}$ = 0.13 $\pm$ 0.05 MeV \cite{davis2005}}. 
Several theoretical calculations predict a value of the $^3_{\Lambda}$H lifetime compatible with the free $\Lambda$ lifetime \cite{kamada1998,dalitz1966,congleton1992,gal2018}. However, recent results from heavy-ion experiments are below the expectations, leading to the so called ``lifetime puzzle".

The ALICE experiment measured the $^3_{\Lambda}$H lifetime with higher precision using the data sample of Pb--Pb collisions at $\sqrt{s_{\mathrm{NN}}}$ = 5.02 TeV \cite{alice_3lh_2019}. The $^3_{\Lambda}$H is reconstructed and selected with the same approach used for production yields.
The method adopted for the lifetime determination is the exponential fit to the corrected d$N$/d$ct$ spectrum which leads to a value of \mbox{$\tau$ = 242$^{+34}_{-38}$ (stat.)} $\pm$ 17 (syst.) ps. A second analysis technique has been used to cross-check the result and consists in an unbinned fit to the 2-dimensional distribution $ct$ vs invariant mass, as described in \cite{ALICE-PUBLIC-2019-003}. The lifetime value obtained with the second method is in agreement, within statistical and systematic uncertainties, with that obtained using the exponential-fit method.

\begin{figure}[h]
	\centering
    \includegraphics[width=\textwidth]{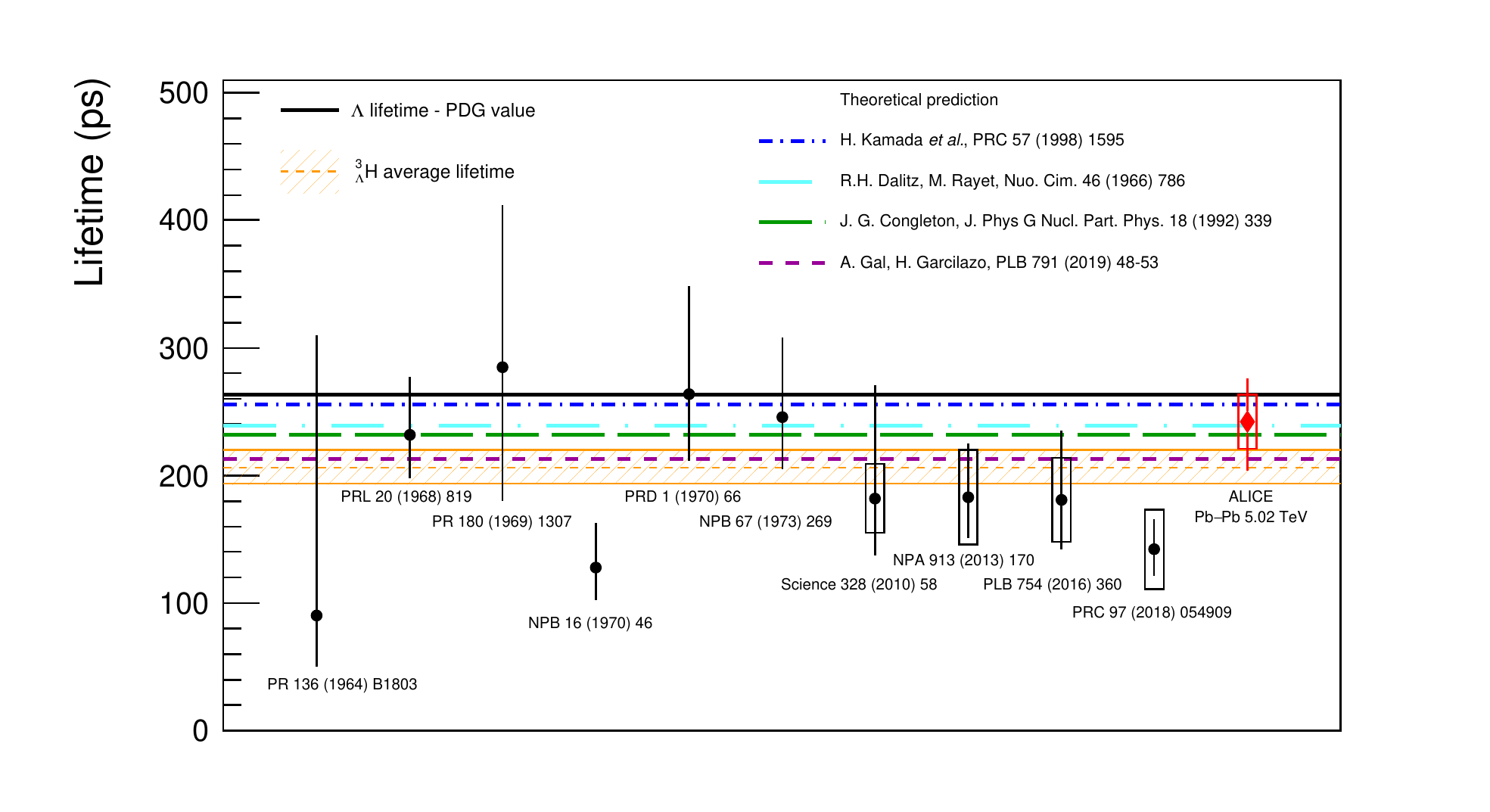}
	\caption{Collection of the $^3_{\Lambda}$H lifetime values obtained with different experimental techniques. The orange band represents the average of the lifetime values and the edges corresponds to 1$\sigma$ uncertainty. The four lines corresponds to theoretical predictions.}
    \label{fig_5}
\end{figure}

Fig.~\ref{fig_5} shows the collection of the lifetime measurements, including the most recent from ALICE in Pb--Pb collisions at $\sqrt{s_{\mathrm{NN}}}$ = 5.02 TeV (red marker), together with representative theoretical calculations. The world average has been computed following the method described in \cite{alice_3lh_2016} and the value obtained is $\tau$ = 206$^{+15}_{-13}$ ps.
 The new result, which at the moment is the one with the highest precision, is in agreement with the predictions. Moreover, it is also compatible with the free $\Lambda$ lifetime, within its statistical uncertainty only.
 
\section{Conclusions}
The new results at the current LHC top energies confirm that both thermal and coalescence models can describe particular aspects of (hyper-)nuclei production in \mbox{heavy-ion} collisions. Nevertheless, the understanding of this picture is evolving thanks to the more precise measurements of the ALICE Collaboration. Furthermore, the measurements of \mbox{(anti-)(hyper-)nuclei} in small collision systems, like pp and p--Pb, hint to a possible unified description of nucleosynthesis, depending only on the system size, and a huge experimental effort is going on to provide more results and shed new light on this topic.

In addition, the $^3_{\Lambda}$H lifetime collection has been enriched with a new and currently more precise value, which is closer to the free $\Lambda$ lifetime and hence compatible with the current theoretical picture. This result goes in the direction of solving the $^3_{\Lambda}$H ``lifetime puzzle" but it also needs to be reinforced with more measurements. Further improvements on the lifetime measurements are expected by exploiting the $^3_{\Lambda}$H decay mode into three bodies. In the future LHC Run 3 and 4, ALICE is expected to collect a larger data sample and improve the results thanks to the upgrade of its detectors. This should allow for a reduction of the statistical uncertainty on the $^3_{\Lambda}$H lifetime to the level of 5$\%$.

\end{document}